\documentclass[5p,preprint,sort&compress]{elsarticle}
\usepackage{graphicx,amsmath,amssymb}

\journal{arXiv}

\begin{document}

\begin{frontmatter}
\title{
  Parametric Resonance in Neutrino Oscillation:
  A Guide to Control the Effects of Inhomogeneous Matter Density
}

\author[uu1]{Masafumi Koike}
\ead{koike@is.utsunomiya-u.ac.jp}
\author[su]{Toshihiko Ota\corref{cor}}
\cortext[cor]{Corresponding author.}
\ead{toshi@mail.saitama-u.ac.jp}
\author[uu2]{Masako Saito}
\ead{m\_saito@cc.utsunomiya-u.ac.jp}
\author[su]{Joe Sato}
\ead{joe@phy.saitama-u.ac.jp}

\address[uu1]{%
  Graduate School of Engineering, Utsunomiya University,
  7-1-2 Yoto, Utsunomiya, Tochigi 321-8585, Japan
}
\address[su]{%
  Physics Department, Saitama University,
  255 Shimo-Okubo, Sakura-ku, Saitama, Saitama 338-8570, Japan
}
\address[uu2]{%
  Faculty of Engineering, Utsunomiya University,
  7-1-2 Yoto, Utsunomiya, Tochigi 321-8585, Japan
}

\date{\today}

\begin{abstract}
  Effects of the inhomogeneous matter density on the three-generation
  neutrino oscillation probability are analyzed.
  Realistic profile of the matter density is expanded into a Fourier series.
  Taking in the Fourier modes one by one, we demonstrate that each mode 
  has its corresponding target energy. 
  The high Fourier mode selectively modifies the oscillation probability 
  of the low-energy region. 
  This rule is well described by the parametric resonance 
  between the neutrino oscillation and the matter effect. 
  The Fourier analysis gives a simple guideline to systematically control 
  the uncertainty of the oscillation probability caused by the uncertain density of matter. 
  Precise analysis of the oscillation probability down to the low-energy region 
  requires accurate evaluation of the Fourier coefficients of the matter density
  up to the corresponding high modes. 
\end{abstract}

\begin{keyword}
  Neutrino oscillation \sep
  Matter effect \sep
  Parametric resonance \sep
  Fourier analysis

  \PACS 14.60.Pq \sep 13.15.+g \sep 14.60.Lm

  STUPP-16-226
\end{keyword}

\end{frontmatter}

\section{Introduction}
Revolutionary discoveries in neutrino experiments over the last two
decades have established the oscillation among the neutrino
flavors~\cite{Fukuda:1998mi,Ahmad:2002jz} and narrowed the allowed
value of oscillation parameters~\cite{globalfit}. 
One of the focal interests in the coming years is the possible
presence of the leptonic CP violation.
While the CP violation gives no more than subleading effects in the
neutrino oscillation, it potentially gives a hint to the origin of the
dominance of the matter over antimatter in the universe~\cite{Leptogenesis}.
Experiments aiming for this subleading effect shall involve a very long
baseline length comparable with the diameter of the Earth.
Neutrinos travel through deep inside the Earth so that the matter
density on the path is sufficiently large to have a significant
role in the oscillation probability~\cite{Wolfenstein:1977ue}.
Control over the effects of the matter density is thus of utmost
importance for determining the oscillation parameters including the
CP-violating phase~\cite{Arafune:1996bt}.

The matter density in the interior of the Earth is evaluated 
by a geophysical Earth model 
such as the Preliminary Reference Earth Model 
(PREM)~\cite{Dziewonski:1981xy}. 
It has a layer structure, which consists of the crust, mantle and core, 
and gives the spatial variation of the density on the neutrino path. 
The evaluated density is subject to uncertainty, due to, for
example, limitations of modeling 
(see Ref.~\cite{Earthmodels} for variety of Earth models)
and the insufficient
knowledge on the chemical composition 
in the deep Earth (see Ref.~\cite{Rott:2015kwa} for recent discussions). 
One should control the uncertainty so as not to obstruct the experimental
search for the oscillation parameters 
(for parameter correlation and uncertainty from matter density profile, 
see e.g. Ref.~\cite{Ota:2002fu}).
Elucidating the required accuracy of the matter density 
for the experimental precision goal   
will enrich the future prospect for the precision measurements. 

From theoretical points of view, 
effects of the matter density on the oscillation probabilities 
have been quantitatively surveyed with plenty of 
numerical
simulations (see e.g. Ref.~\cite{1986:Ermilova,ParametricResonance,Petcov,early:matter,early:simulation,Koike:1998hy,Ota:2000hf,1990s:profile,Koike:2009xf}).
The constant-density models give good approximation for the
experiments with conventional baselines of a few hundred kilometers,
and have been studied most widely.  The impact of the uncertainty of
the density has also been studied for these models~(for early works,
see e.g. Ref.~\cite{early:simulation}).
Recent interests in the very long baseline experiments motivate to
allow for the inhomogeneity of the matter density~%
\cite{1986:Ermilova,ParametricResonance,Petcov,Koike:1998hy,Ota:2000hf,1990s:profile,Koike:2009xf,Akhmedov:2000js}.

Effects of the inhomogeneous matter density are investigated 
by simplified matter profile models.
One of the models is the three-segment (mantle-core-mantle) 
model (see e.g. Ref.~\cite{3layer}).
This model allows easy handling in numerical simulations
and well reproduces oscillation probabilities 
calculated with the full Earth model.
We propose a complimentary model in this study: the Fourier-series model, which
consists of first few Fourier modes evaluated from the Earth
model~\cite{Koike:1998hy,Ota:2000hf,Koike:2009xf}.
This model admits the systematic improvement of the matter density
profile and thus of the oscillation probability. 

This paper is organized as follows. 
In Sec.~\ref{sec:param-res-2gen}, 
we recapitulate our previous study of the Fourier analysis of the
parametric resonance in the neutrino oscillation~\cite{Koike:2009xf}. 
The parametric resonance takes place when 
the matter profile has the Fourier mode of the matching frequency. 
The study provides a framework for the matter effect. 
In Sec.~\ref{sec:Fourier-analysis}, 
we calculate numerically the oscillation probabilities for 
the matter profile of the full PREM and 
the approximated profiles with the first few Fourier modes. 
We study in Sec.~\ref{sec:param-res-3gen} the improvement of the approximation 
by the higher Fourier mode in terms of the parametric resonance. 
Conclusion and discussions are given in Sec.~\ref{sec:conclusion}. 

\section{Parametric resonance in the Two-generation Neutrino Oscillation}
\label{sec:param-res-2gen}
We give a short review of the matter effect on the neutrino oscillation 
and parametric resonance presented in Ref.~ \cite{Koike:2009xf}. 
See this reference for the full discussion.

We employ the two-generation neutrino oscillation 
between $\nu_{\mathrm{e}}$ and $\nu_{\mu}$ driven by the evolution equation of
\begin{subequations}
\begin{gather}
  \mathrm{i} \frac{\mathrm{d}}{\mathrm{d} x}
  \begin{pmatrix} \nu_{\mathrm{e}}(x) \\ \nu_{\mu}(x) \end{pmatrix}
  =
  H(x) 
  \begin{pmatrix} \nu_{\mathrm{e}}(x) \\ \nu_{\mu}(x) \end{pmatrix}
 \label{eq:evolution_Diffeq_3}%
 \intertext{with}
  H(x)
  =
  \frac{1}{4E}
  \begin{pmatrix}
    - \delta m^{2}\cos 2\theta + 2 a(x) & \delta m^{2}\sin 2\theta \\ 
      \delta m^{2}\sin 2\theta & \delta m^{2}\cos 2\theta
  \end{pmatrix}
  \, .
 \label{eq:evolution_H_3}%
\end{gather}%
\label{eq:evolution_eq_3}%
\end{subequations}
Here
$E$ is the energy,  
$\delta m^{2}$ is the quadratic mass difference of the
neutrinos, $\theta$ is the mixing angle, and
\begin{math}
 a(x) \equiv 
 2\sqrt{2}
 E G_{\textrm{F}} N_{\textrm{A}} Y_{\textrm{e}}
 \rho(x) / [1.0 \, \mathrm{g/mol}]
\end{math}
is the matter effect under the matter density of $\rho(x)$ 
where $G_{\textrm{F}}$, $N_{\textrm{A}}$, and $Y_{\textrm{e}}$ are the 
Fermi constant, the Avogadro constant, and the proton-to-nucleon 
ratio, respectively.
The matter density on the baseline is expanded into Fourier cosine series as 
\begin{equation}
  \rho(\xi) = \rho_{0} + \sum_{n = 1}^{\infty} \rho_{n} \cos 2n \pi \xi \, ,
  \label{eq:rho_n}
\end{equation}
where $\xi = x/L$ with $L$ being the baseline length. 
The matter effect $a(x)$ is accordingly expanded as
\begin{equation}
  a(\xi) =  a_{0} + \sum_{n = 1}^{\infty} a_{n} \cos 2n \pi \xi\, .
  \label{eq:a_n}  
\end{equation}

We assume a simplified model which consists only of the $n$-th Fourier mode of 
the matter effect $a_{n}$ around the average value $a_{0}$ as
\begin{equation}
  a(\xi) =  a_{0} + a_{n} \cos 2n \pi \xi\, . 
\end{equation}
%
%
The evolution equation of neutrino is then reduced to
\begin{align}
  z''(\xi) +
  \bigl( \omega^{2} + U(\xi) \bigr)
  z(\xi) = 0 \, ,
\label{eq:diff-eq-z}
\end{align}
where
\begin{equation}
   z(\xi) =
   \nu_{\textrm{e}}(\xi)
   \exp \Bigl(
     \frac{\mathrm{i}L}{4E} \int_{0}^{\xi} \mathrm{d}s \, a(s)
   \Bigr). 
\end{equation}
Here we assume that the initial state of neutrino is muon neutrino,
giving the initial condition of $\nu_{\textrm{e}}(0) = 0$ and
$\nu_{\mu}(0) = 1$.
Equation \eqref{eq:diff-eq-z} is a simple harmonic oscillator equation 
perturbed by the inhomogeneity of the matter effect. 
The appearance probability of electron neutrino $P(\nu_{\mu} \to \nu_{\textrm{e}})$ 
is given by 
$P(\nu_{\mu} \to \nu_{\textrm{e}}) = \vert \nu_{\textrm{e}}(1) \vert^{2} = \vert z(1) \vert^{2}$. 
The natural frequency of the system in Eq.~(\ref{eq:diff-eq-z}) 
\footnote{%
  The definition of $\omega$ in Eq.~(\ref{eq:diff-eq-z-2}) is 
  different from Eq.~(6b) of Ref.~\cite{Koike:2009xf} by the term of $D$.
  In this paper, we include this term into $U(\xi)$ rather than into $\omega$. %
}
, given by 
\begin{equation}
  \omega
  =
  \frac{L}{4E}
  \bigl[
    ( a_{0} - \delta m^{2} \cos 2\theta )^{2}
    + ( \delta m^{2} \sin 2\theta )^{2}
  \bigr]^{1/2},
\label{eq:diff-eq-z-2}%
\end{equation}
is controlled by the average matter effect $a_{0}$. 
On the other hand, the perturbation $U(\xi)$ is 
led by the $n$-th Fourier mode of matter effect $a_{n}$ as seen in
\begin{equation}
  U(\xi) =
   \alpha_{n} \cos 2n\pi\xi
   - \mathrm{i} \beta_{n} \sin 2n\pi\xi
   + \gamma_{n} \cos^{2} 2n\pi\xi,
\end{equation}
where
\begin{align}
   \alpha_{n} &= 
   \frac{a_{n}L}{4E} \Bigl( \frac{a_{0}L}{2E} -  \frac{\delta m^{2} L}{2E} \Bigr),
   \\
   \beta_{n} &=
   n \pi \frac{a_{n}L}{2E},
   \\
   \gamma_{n} &=
   \Bigl( \frac{a_{n}L}{4E} \Bigr)^{2}
   \, .
\end{align}

Equation~(\ref{eq:diff-eq-z}) is a Hill's equation~\cite{HillEquation}, 
which is a generalization of 
Mathieu's equation that appears in typical parametric resonance 
problems. 
It leads to the parametric resonance by the perturbation 
when its frequency $2n\pi$ matches twice of the natural frequency $2\omega$. 
This resonance condition reads $E = E_{n}^{(\pm)}$ with 
\begin{equation}
  E_{n}^{(\pm)} =
  \frac{ \delta m^{2} E}{
    a_{0} \cos 2\theta
    \pm \sqrt{
      \bigl( 4 n \pi E/L \bigr)^{2}
      - a_{0}^{2} \sin^{2} 2\theta
    }
  } \, ,
\label{eq:parametric_resonance_energy}
\end{equation}
where $E_{n}^{(+)}$ and $E_{n}^{(-)}$ lie below and above 
the MSW resonance energy $E_{\textrm{MSW}}$ given by
\begin{equation}
  E_{\textrm{MSW}} \equiv
  \frac{ \delta m^{2} E \cos 2\theta}
       {a_{0}} \, .
\label{eq:MSW_energy}
\end{equation}
We have shown in Ref.~\cite{Koike:2009xf} that 
this analogy of the parametric resonance indeed works in the two-generation analysis. 
The Fourier mode of the matter inhomogeneity takes effects upon 
the oscillation probability 
around the corresponding resonance energy $E_{n}^{(\pm)}$. 
Note that the oscillation probability reaches minimum at the resonance
energies since $\omega = n\pi$ holds at $E = E_{n}^{(\pm)}$.

\section{Fourier Analysis of the Matter Effect on the Oscillogram}
\label{sec:Fourier-analysis}

We investigate the effect of the inhomogeneous matter density 
under more realistic condition. 
We refine our analysis in the following two aspects.
First, we incorporate the third generation of neutrinos.  Two
quadratic mass differences of $\delta m_{21}^{2}$ and $\delta
m_{31}^{2}$ are allowed in this formalism.  The smaller difference
$\delta m_{21}^{2}$ which was neglected in the two-generation analysis
becomes dominant in the low-energy region of $E \lesssim 1 \,
\mathrm{GeV}$.  Second, we use a realistic matter density profile
estimated from internal structure of the Earth.  We employ the
Preliminary Reference Earth Model (PREM) as an Earth
model~\cite{Dziewonski:1981xy} in the analysis.  We expand the matter
profile of the PREM into a Fourier series and investigate the effect
of each Fourier mode.  We take in more than one Fourier mode and
include mode-mode coupling effect, which we omitted in the analysis in
the previous section.
We carry out the calculation numerically and present the results by
the oscillogram, which is a contour plot of the oscillation probability 
on the plane of neutrino energy and baseline length.

\begin{figure}%
  \includegraphics[width=80mm]{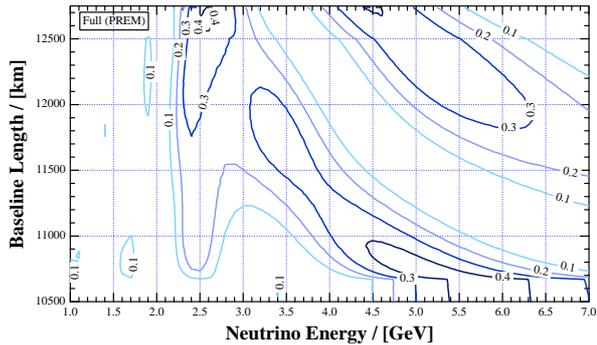}
  \caption{
    The oscillogram for the matter profile by the PREM. 
    Oscillation parameters are taken as in the text.
    }
  \label{fig:Pme_Full}
\end{figure}

We show in Fig.~\ref{fig:Pme_Full} an oscillogram of 
the appearance probability $P_{\mathrm{PREM}}(\nu_{\mu} \to \nu_{\textrm{e}})$ 
assuming the matter profile by the PREM. 
In our plot, the value of the neutrino energy $E$ and that of the
baseline length $L$ range over $(1 \textrm{---} 7)\, \mathrm{GeV}$
and $(10500 \textrm{---} 12700)\, \mathrm{km}$, respectively.
We take the values of oscillation parameters in this section as
follows~\cite{globalfit,Agashe:2014kda}:
\begin{equation}
\begin{gathered}
 \delta m_{21}^{2} = 7.45 \cdot 10^{-5} \, \mathrm{eV^{2}},
 \,
 \delta m_{31}^{2} = 2.42 \cdot 10^{-3} \, \mathrm{eV^{2}},
 \\
 \sin \theta_{12} = 0.553,
 \,
 \sin \theta_{23} = 0.668,
 \,
 \sin \theta_{13} = 0.152,
 \\
 \delta_{\textrm{CP}} = \pi/2.
\end{gathered}%
 \label{eq:oscgram-params}
\end{equation}%
The major peak 
in Fig.~\ref{fig:Pme_Full} extends diagonally from
$(E, L) \simeq (2.5 \, \mathrm{GeV}, 12700 \, \mathrm{km})$ to
$(6 \, \mathrm{GeV}, 10700 \, \mathrm{km})$, and accompanies 
additional peaks alongside.
The overall pattern of the contour plot is not simple, 
because the matter density profile significantly depends on the baseline length. 
The contours show clear cusps at the length $L \simeq 10700 \, \mathrm{km}$, 
beyond which the baseline cut the mantle-core boundary of the internal structure 
of the Earth. 
Other cusps observed at $L \simeq 12500\, \mathrm{km}$  
correspond to the boundary of the outer-inner core. 

\begin{figure}
  \includegraphics[width=80mm]{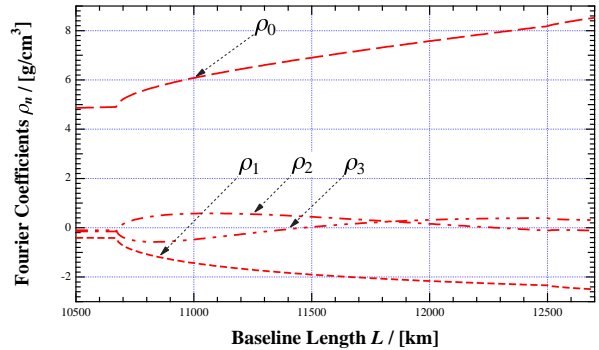}
  \caption{
    Average matter density $\rho_{0}$ and 
    the first three Fourier coefficients $\rho_{1}$, $\rho_{2}$ and $\rho_{3}$
    as functions of baseline length $L$.
    The density profile function for each $L$ is generated based 
    on the PREM. 
    }
  \label{fig:DensityCoeffs}  
\end{figure}

We next systematize the overall pattern of the oscillogram 
with the help of the Fourier analysis of the matter effect.
The matter density profile on a baseline is symmetric due to the
spherical symmetry of the PREM.  The matter effect can be expanded in
a Fourier cosine series as in Eq.~(\ref{eq:a_n}).  Figure
\ref{fig:DensityCoeffs} shows the values of the first four Fourier
coefficients as functions of the baseline length $L$.  The curves have
kinks at $L \simeq 10700\, \mathrm{km}$ and $L \simeq 12500\,
\mathrm{km}$, reflecting the internal structure of the Earth mentioned
above.  We are able to construct systematic approximations of the
profile by truncating the series after these first few modes.

\begin{figure}
  \includegraphics[width=80mm]{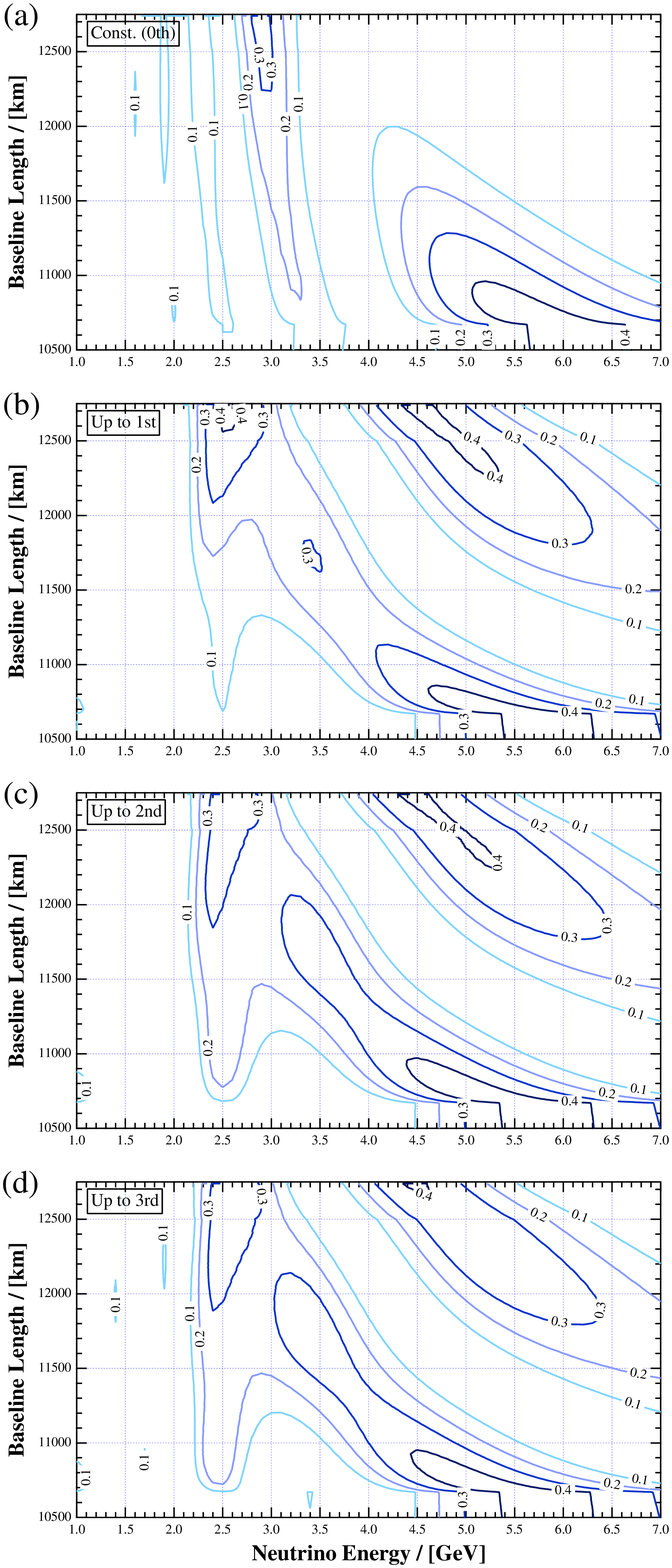}
  \caption{
    Oscillograms calculated with the matter profile taken
    up to the (a) zeroth, (b) first, (c) second, and (d) third Fourier modes.
    }
  \label{fig:Pme_All}  
\end{figure}

We draw oscillograms for four models of matter profiles. 
We start with the model with only the constant matter density (the zeroth Fourier mode) 
and take in the Fourier mode one by one to improve the matter profile systematically~\cite{Ota:2000hf}.
Figure~\ref{fig:Pme_All} presents the oscillograms calculated with the matter profile 
taken up to the zeroth, first, second, and third Fourier modes. 
The values of each Fourier coefficient are taken from Fig. \ref{fig:DensityCoeffs}.
The oscillogram for the constant matter density displayed in
Fig.~\ref{fig:Pme_All}~(a) shows the major peak
that diagonally runs from
$(E, L) \simeq (4 \, \mathrm{GeV}, 12000 \, \mathrm{km})$
to $(6 \, \mathrm{GeV}, 10700 \, \mathrm{km})$ and narrow peaks
aside it. 
The oscillogram deviates qualitatively from Fig.~\ref{fig:Pme_Full} for the PREM. 
The deviation becomes smaller in Fig.~\ref{fig:Pme_All}~(b) 
and is contained only in the low-energy region. 
Oscillograms of Figs.~\ref{fig:Pme_All}~(c) and (d) 
reproduce almost perfectly 
the PREM oscillogram of Fig.~\ref{fig:Pme_Full} in all energy region.
This result is consistent with our previous study in the
two-generation model~\cite{Koike:2009xf}: each $n$-th Fourier mode of
the matter profile selectively modifies the oscillation probability at
around the corresponding energy, which we call the parametric
resonance energy and is given by $E_{n}^{(\pm)}$ of
Eq.~(\ref{eq:parametric_resonance_energy}).
We will take a closer look at consistency
between the numerical results shown with the oscillograms 
and the analytic formula of the oscillation probabilities with 
Fourier-expanded density profile in the next section.

\section{Parametric Resonance in the Three-generation Neutrino Oscillation}
\label{sec:param-res-3gen}
\begin{figure}
  \includegraphics[width=80mm]{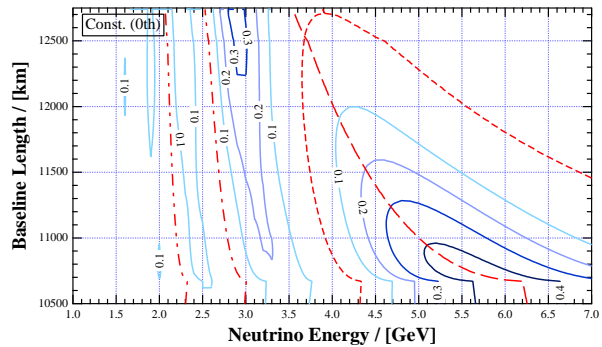}
  \caption{
    The oscillogram for the constant density model
    (Fig.~\ref{fig:Pme_All}~(a)), overlaid by the resonance energies.
    The dashed, dotted, chain, and two-dot chain red curves represent
    $E_{\mathrm{MSW}}$, $E_{1}^{(\pm)}$, $E_{2}^{(+)}$ and
    $E_{3}^{(+)}$, respectively.
  }
  \label{fig:Pme_Rho0_Resonance} 
\end{figure}

We show that the analogy of the parametric resonance still holds in
the three-generation neutrino oscillation.
It is to be confirmed that the energy region of $E \simeq
E_{n}^{(\pm)}$ in the oscillogram is sensitive to the $n$-th Fourier
mode of the matter profile.  First we determine the two-generation
parameters in terms of the three-generation ones.
Under the present setup, the mass difference $\delta m_{31}^{2}$ and
the matter effect are essential to the oscillation.  We thus interpret
the two-generation parameters as
\begin{equation}
\begin{gathered}
  \delta m^{2} \to \delta m^{2}_{31} = 2.42 \cdot 10^{-3} \, \mathrm{eV^{2}}, \\
  \sin \theta \to \sin \theta_{13} = 0.152 \, .
  \label{eq:2gen-3gen-corresp}
\end{gathered}  
\end{equation}
The values of the parametric resonance energy $E_{n}^{(\pm)}$
for Eq.~(\ref{eq:2gen-3gen-corresp}) 
are listed in Table \ref{table:PREM-Eres}. 
The MSW resonance energy of Eq.~(\ref{eq:MSW_energy}) is also listed. 
\begin{table}
  \caption{
    The resonance energies $E_{n}^{(\pm)}$ 
    for $n=1$, $2$ and $3$ and $E_{\mathrm{MSW}}$.
    The baseline length is taken as 
    $L = 10500$, $11000$, $11500$, $12000$ and $12500\, \mathrm{km}$.
    }
    \begin{tabular}{cccccc}
      \hline\hline
      $\dfrac{L}{\mathrm{[km]}}$
      & $\dfrac{E_{3}^{(+)}}{\mathrm{[GeV]}}$
      & $\dfrac{E_{2}^{(+)}}{\mathrm{[GeV]}}$
      & $\dfrac{E_{1}^{(+)}}{\mathrm{[GeV]}}$
      & $\dfrac{E_{\mathrm{MSW}}}{\mathrm{[GeV]}}$
      & $\dfrac{E_{1}^{(-)}}{\mathrm{[GeV]}}$
      \\
      \hline
      $10\,500$ & $2.30$ & $2.99$ & $4.32$ & $6.25$ & $16.81$ \\
      $11\,000$ & $2.20$ & $2.79$ & $3.92$ & $5.01$ & $9.22$ \\
      $11\,500$ & $2.15$ & $2.69$ & $3.74$ & $4.41$ & $6.86$ \\
      $12\,000$ & $2.12$ & $2.62$ & $3.66$ & $4.01$ & $5.55$ \\
      $12\,500$ & $2.09$ & $2.57$ & $3.67$ & $3.72$ & $4.60$ \\
      \hline\hline
    \end{tabular}
  \label{table:PREM-Eres}
\end{table}
We plot the resonance energies of $E_{n}^{(\pm)}$ and $E_{\mathrm{MSW}}$ 
in Fig.~\ref{fig:Pme_Rho0_Resonance} by the red curves, 
laid over the oscillogram for the constant density 
of Fig.~\ref{fig:Pme_All}~(a). 
The energy $E_{n}^{(\pm)}$ runs along the dip of the probability, 
while $E_{\mathrm{MSW}}$ lies just on the major peak in the oscillogram 
as in the two-generation spectrum. 
The consequence summarized in Fig.~\ref{fig:Pme_Rho0_Resonance} allows
us to understand the oscillograms with the help of the two-generation
model.

\begin{figure}
  \includegraphics[width=80mm]{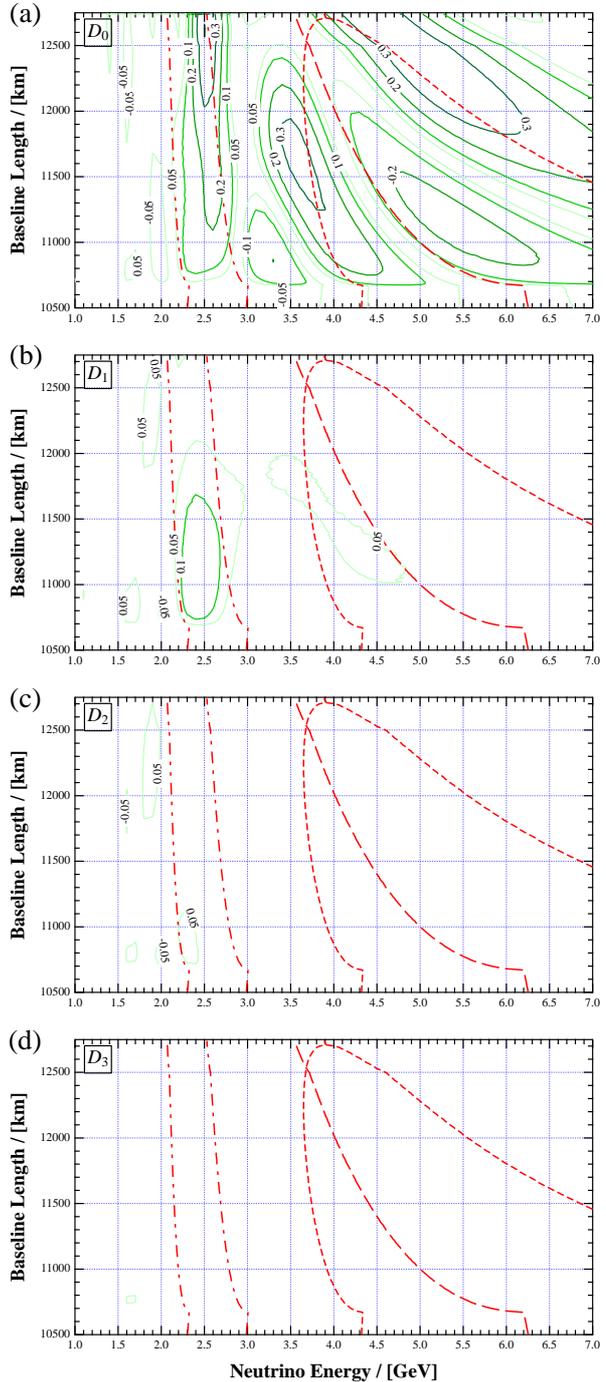}
  \caption{
    Residual difference 
    between the appearance probabilities for the PREM
    and for the matter profile with up to the $n$-th Fourier mode.
    The green solid curves represent the residual difference
    defined by Eq.~(\ref{eq:residual_diff}) for (a) $n = 0$, (b)
    $n = 1$, (c) $n = 2$, and (d) $n = 3$.
    The overlaid dashed, dotted, chain, and two-dot chain red curves
    are same as in Fig.~\ref{fig:Pme_Rho0_Resonance}.
  }
  \label{fig:dPme_All}
\end{figure}

We closely study 
the disagreement between the oscillograms 
of Figs.~\ref{fig:Pme_Full} and \ref{fig:Pme_All}, 
focusing to the resonance energies $E_{n}^{(\pm)}$ and $E_{\mathrm{MSW}}$. 
We define 
\begin{equation}
 D_{n}
 \equiv
 P_{\textrm{PREM}}(\nu_{\mu} \to \nu_{\textrm{e}})
 -
 P_{n}(\nu_{\mu} \to \nu_{\textrm{e}}),
\label{eq:residual_diff} 
\end{equation}
which is the residual difference 
between the appearance probability for the PREM 
$P_{\textrm{PREM}}(\nu_{\mu} \to \nu_{\textrm{e}})$
and that for the matter profile taken up to the $n$-th Fourier mode 
$P_{n}(\nu_{\mu} \to \nu_{\textrm{e}})$.
This value represents the error of the approximation up to the
$n$-th Fourier mode.
In Fig.~\ref{fig:dPme_All} 
the values of $D_{n}$ for $n = 0$, $1$, $2$, and $3$ 
are shown by the green curves.
The overlaid red curves are same as in Fig.~\ref{fig:Pme_Rho0_Resonance}. 

The value of $D_{0}$ is found  
to remain sizable in the whole energy region in Fig.~\ref{fig:dPme_All}~(a). 
This signifies that the disagreement between the oscillogram for the
PREM (Fig.~\ref{fig:Pme_Full}) and that for the constant density model
(Fig.~\ref{fig:Pme_All}~(a)) extends over the whole energy region.
Juxtaposing Fig.~\ref{fig:dPme_All}~(b) against Fig.~\ref{fig:dPme_All}~(a), we find
the remarkable reduction of 
the residual difference at around the red dotted curve of $E_{1}^{(\pm)}$. 
The persistent difference $D_{1}$ is thereby restricted 
in the low-energy region. 
Similarly, the value of $D_{2}$ in Fig.~\ref{fig:dPme_All}~(c) 
diminishes around the curve of $E_{2}^{(+)}$, 
while Figure ~\ref{fig:dPme_All}~(d) clears the difference around $E_{3}^{(+)}$. 
In this manner, 
the improvement of the oscillogram by the $n$-th Fourier mode is systematically 
controlled by the resonance energy $E_{n}^{(\pm)}$. 

The comparison between oscillograms allows us to quantitatively analyze 
impacts of the higher Fourier modes.
For example, the matter profile with the Fourier modes up to $n=2$ 
reproduces the oscillation probability calculated for the PREM profile 
with high accuracy in all the baseline region
with $E > 2.5$ GeV, as seen in Fig.~\ref{fig:dPme_All} (c). 
The inclusion of three Fourier coefficients  
$a_{0}, a_{1}$, and $a_{2}$ is thus enough 
to simulate accurately an experiment whose threshold energy for detecting neutrinos 
is larger than 2.5 GeV. 

We conclude that each Fourier mode of the matter profile has 
its corresponding target energy: 
the effect of the $n$-th mode appears in the oscillogram 
at around the resonance energy $E_{n}^{(\pm)}$. 
This simple correspondence remains valid 
in realistic three-generation neutrino oscillations.

\section{Conclusion and Discussions}
\label{sec:conclusion}

We investigated effects of the inhomogeneous matter density 
on the long baseline neutrino oscillation.
We evaluated the appearance probability $P(\nu_{\mu} \to \nu_{\textrm{e}})$ 
and showed it in the oscillograms, 
which are contour plots on the plane of the neutrino energy and the baseline length. 
The baseline length is assumed to be long enough to go through 
the core of the Earth. 
The matter density on the baseline varies according to  
the internal structure of the Earth, which we evaluate 
using the Preliminary Reference Earth Model (PREM). 

We expanded the matter density profile into a Fourier series and truncated it 
to obtain four density profiles: constant density profile and the profiles 
taken up to the first, second, and third Fourier modes.
We drew the oscillograms for the profiles and compared them to that for the full PREM.
As we append the higher Fourier modes to the density profile, 
the oscillograms better reproduce the one for the full PREM.
We clarified the systematics behind the improvement with the aid of 
the parametric resonance: the $n$-th Fourier mode selectively modifies 
the oscillogram at around the parametric resonance energy $E_{n}^{(\pm)}$. 
The analogy of the parametric resonance, 
which we derived in the two-generation framework~\cite{Koike:2009xf}, 
is demonstrated to remain valid in realistic three-generation neutrino oscillations.

Our study elucidates the correspondence 
between the length scale of matter density variation and 
the energy of neutrinos. 
The high Fourier mode, 
which represents the small-scale variation of the matter density,  
is shown to have significance for the oscillation of the low-energy neutrinos. 
The correspondence is clear when we note  
that the low-energy neutrino has the short oscillation length 
and is sensitive to the small-scale variation. 
Our result is consistent to the studies on the Earth tomography with
neutrinos~\cite{Tomography}.  It is shown that the low-energy
neutrinos carry the information on the fine structure of the internal
Earth on the baseline.  The fine structure in our analysis is given by
the high Fourier mode, which improves the oscillation probability down
to the corresponding resonance energy.

The Fourier analysis gives a simple guideline to systematically control 
the uncertainty of the oscillation probability caused by the uncertain density of matter. 
Precise analysis of the oscillation probability down to the low-energy region 
requires accurate evaluation of the Fourier coefficients of the matter density
up to the corresponding high modes. 
We are planning to discuss in the next study 
the correlation between the Fourier coefficients of the matter density 
and the oscillation parameters. 
The impact of the uncertainty of the Fourier coefficients will be also studied.

\section*{Acknowledgements}

This work was supported by MEXT Grants-in-Aid for Scientific
Research (KAKENHI) Grant No. 25105009.
This work was supported also by JSPS Grants-in-Aid for Scientific
Research (KAKENHI) Grant No. 24740145 (M.K.), No. 26105503 (T.O.), and
No. 24340044 (J.S.).


\end{document}